\documentclass[pre,twocolumn,aps,superscriptaddress,showpacs,floatfix]{revtex4}
\usepackage{amsmath}
\usepackage{graphicx}

\begin{document}
\title{Vibrational ratchets}
\author{M. Borromeo}
\affiliation{Dipartimento di Fisica, Universit\`a di Perugia,
I-06123 Perugia, Italy}
\affiliation{Istituto Nazionale di Fisica Nucleare, Sezione di
Perugia, I-06123 Perugia, Italy}
\author{F. Marchesoni}
\affiliation{Dipartimento di Fisica, Universit\`a di Camerino,
I-62032 Camerino, Italy}
\date{\today}
\begin{abstract}
Transport in one-dimensional symmetric devices can be activated
by the combination of thermal noise and a bi-harmonic drive. For
the study case of an overdamped Brownian particle diffusing on a
periodic one-dimensional substrate, we distinguish two apparently
different bi-harmonic regimes: (i) {\it Harmonic mixing}, where
the two drive frequencies are commensurate and of the order of
some intrinsic relaxation rate. Earlier predictions based on
perturbation expansions
seem inadequate to interpret our simulation results; 
(ii) {\it Vibrational mixing}, where one
harmonic drive component is characterized by high frequency but
finite amplitude-to-frequency ratio. Its effect on the device
response to either a static or a low-frequency additional input
signal is accurately reproduced by rescaling each spatial Fourier
component of the substrate potential, separately. Contrary to
common wisdom, based on the linear response theory, we show that
extremely high-frequency modulations can indeed influence the
response of slowly (or dc) operated devices, with potential
applications in sensor technology and cellular physiology.
Finally, the mixing of two high-frequency beating signal is also
investigated both numerically and analytically.
\end{abstract}
\maketitle

\section{Introduction} \label{S1}

The spectral density of the response of a system at 
thermodynamical equilibrium to a sinusoidal time modulation consists 
of a delta-like spike centered at the forcing frequency. 
In linear response theory, the system response to two sufficiently
weak sinusoidal modulations with different frequency is well reproduced
by the linear superposition of the system response to each
modulation, separately. In other words, the spectral contents 
of the system output coincides with that of the input signal. 
A significant exception is represented by the {\it harmonic mixing}
(HM) of two commensurate input frequencies of comparable magnitude
\cite{seeger}: The system response then can contain harmonics of 
both drive frequencies and thus, under certain conditions, even a dc component.

In this paper we focus on the regime when at least one drive 
component is characterized by high frequency but finite 
amplitude-to-frequency ratio. Such a fast modulation modifies the internal
dynamics of the system, so that its response to the other harmonic drive
is sensitive to both modulating frequencies, no matter what their ratio.
Such a frequency coupling, termed {\it vibrational mixing}, does not
fall within the framework of the linear response theory, as the amplitude 
of at least one drive component must be appreciably large. Right for this
reason, however, the regime investigated here is consistent with the 
operating conditions of many real devices and is, indeed, 
of general applicability.

Our study case is reprensented by a Brownian 
particle moving on a one dimensional
substrate subjected to an external bi-harmonic force $F(t)$ and a
zero-mean valued, delta-correlated Gaussian noise $\xi(t)$. Its
coordinate $x(t)$ obeys the Langevin equation (LE)
\begin{equation}
\label{1.1} \dot x= -V'(x) +F(t) + \xi(t),
\end{equation}
where
\begin{equation}
\label{1.2} F(t) = A_1\cos(\Omega_1 t+\phi_1) +A_2\cos(\Omega_2
t+\phi_2)
\end{equation}
with $A_1, A_2 \geq 0$,
\begin{equation}
\label{1.3} \langle \xi(t) \rangle =0, ~~~ \langle \xi(t) \xi(0)
\rangle =2 D \delta (t),
\end{equation}
and $V(x)$ is the periodic potential of a substrate with period
$L=2\pi$.

In Sec. II we compare the results of extensive numerical
simulations with earlier perturbation predictions for the
rectification current $\langle \dot x \rangle/2\pi$ induced by HM.
We conclude that, in spite of the abundance of numerical results,
the analytical description of HM available in the literature is
still incomplete and, to some extent, unsatisfactory. We
then introduce the vibrational mixing regime. A
high-frequency perturbation pumps energy into the system forcing
free particle oscillations of amplitude $\psi_0 =A_2/\Omega_2$
comparable with the system length-scale.  In Sec. III we
demonstrate both numerically and analytically that the particle
response to an additional dc drive is extremely sensitive to the
high-frequency pump parameter $\psi_0$. In Sec. IV we extend our
approach to investigate the rectification current in a rocked
ratchet driven by a bi-harmonic force with
high and low frequency components (vibrational rocked ratchet).
Finally, in Sec. V we consider the case of a ratchet driven by two
high-frequency beating harmonic forces. We show that in such a
limit a vibrational ratchet can be assimilated to a pulsated
ratchet, where the modulation frequency of the substrate amplitude
corresponds to the drive beating frequency.

\section{Harmonic mixing}\label{S2}

We know from the literature of the 1970's \cite{seeger} that a
charged particle confined onto a nonlinear substrate is capable of
mixing two alternating input electric fields of angular
frequencies $\Omega_1$ and $\Omega_2$; its response is expected to
contain harmonics of $\Omega_1$ and $\Omega_2$. As a
result, for commensurate input frequencies, i.e.,
$m\Omega_1=n\Omega_2$, the time dependent particle velocity would
contain a dc component, too. Such a phenomenon, termed in the
later literature {\it harmonic mixing}, is a rectification
effect induced by the asymmetry of the applied force. In view of
general perturbation arguments, HM was predicted to be of the
$(n+m)$th order in the dynamical parameters of the system
\cite{wonneberger,hm}. Lately, HM was re-interpreted as a
manifestation of the ratchet phenomenon \cite{reimann,flach}, even if no
substrate asymmetry is required to generate a HM
signal.

More recently, the HM mechanism has been investigated numerically
as a tool to control the transport of interacting particles in
artificially engineered quasi-onedimensional channels
\cite{riken1,riken2}. An interesting variation of this problem has
been proposed in the context of soliton dynamics, where the
combination of two ac driving forces was proven to rectify the
motion of a kink-bearing chain owing to the inherent nonlinearity
of a travelling kink \cite{solitons}. 

\begin{figure}[htbp]
\centering
\includegraphics[width=8.8cm]{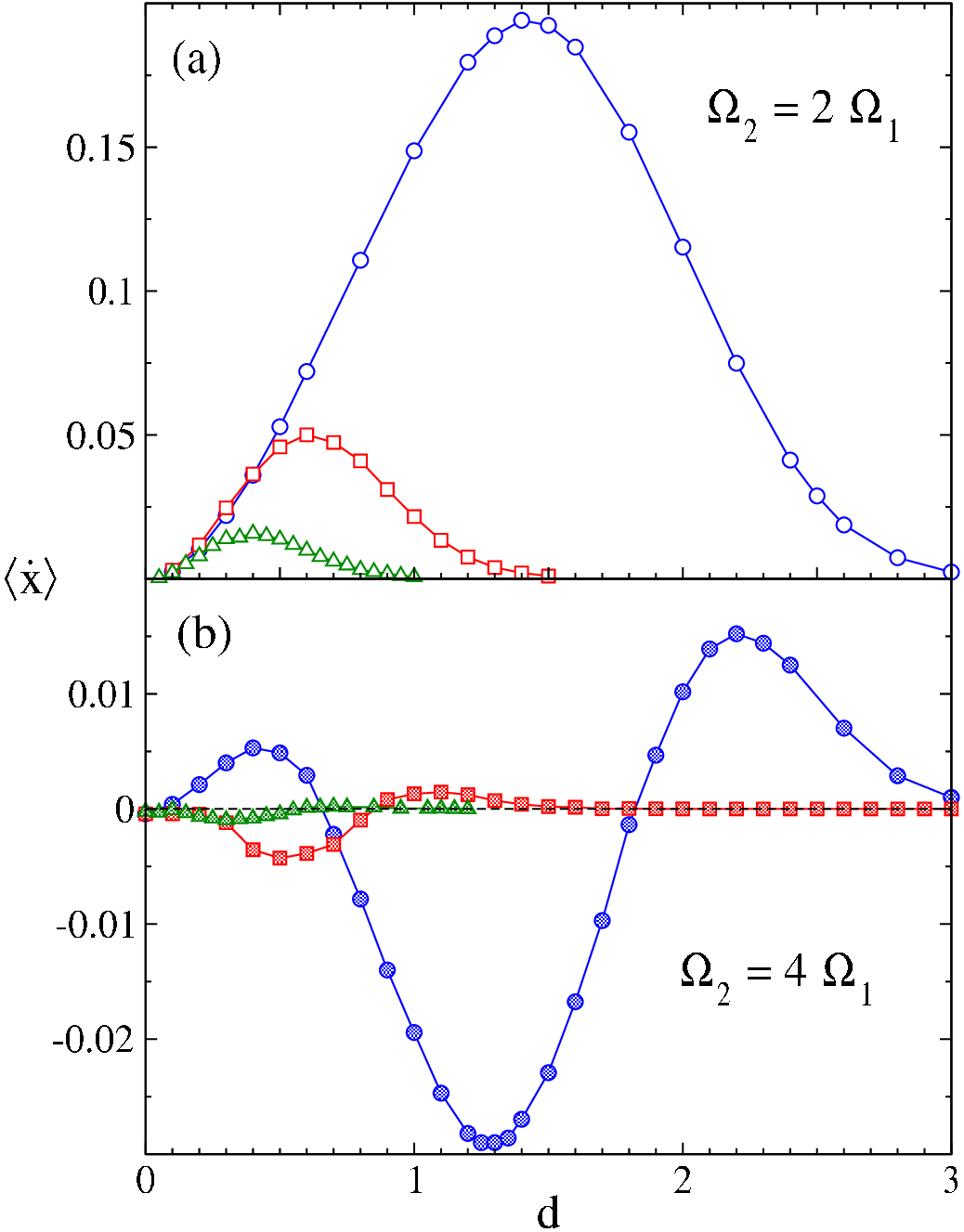}
\caption{(Color online) Transport via HM in the cosine potential (\ref{2.2}) for
$\phi_1=\phi_2$, $A_1=A_2$, and (a) $\Omega_2=2\Omega_1$, (b)
$\Omega_2=4\Omega_1$: $\langle \dot x \rangle$ versus $d$.
Simulation parameters: $\Omega_1= 0.01$, $D=0.2$, and $A_1=0.2$
(triangles), $A_1=0.4$ (squares), and $A_1=1.1$ (circles). }
\label{F1}
\end{figure}

Let us consider, for simplicity, the overdamped stochastic dynamics
(\ref{1.1}) driven by the bi-harmonic force
\begin{equation}
\label{2.1}
F(t)=A_1 \cos (\Omega_1 t) + A_2 \cos(\Omega_2 t)
\end{equation}
with
\begin{equation}
\label{2.2}
V(x) = d(1-\cos x)
\end{equation}
and $\Omega_2=2\Omega_1$. A truncated continued fraction expansion
\cite{wonneberger} led to conclude that in the regime of low
temperature, $d \gg D$, the nonvanishing dc component $\langle
\dot x\rangle$ of the particle velocity would scale like
\begin{equation}
\label{2.3}
\frac{\langle \dot x\rangle}{D} \propto -
\left (\frac{A_1}{2D} \right )^2 \frac{A_2}{2D}.
\end{equation}
Quite surprisingly, this result suggests that for small drive amplitudes and high
substrate barriers, $A_1,A_2 \ll D \ll d$, the HM signal is {\it
negative} and {\it independent} of $d$, at variance with the numerical
results reported in Fig. \ref{F1}. Numerical simulation runs for
increasing $d$ values reveal a resonant $\langle \dot x(d)\rangle$
curve. This is not unexpected as for $d \to 0$ (flattening substrate)
the zero-mean force (\ref{2.1}), with $\langle F(t) \rangle =0$,
cannot sustain a non-null drift current, whereas for $d \to
\infty$ (high substrate barriers) the interwell activation
mechanism gets exponentially suppressed and the relevant drift
current drops to zero. [The conflicting sign in Eq. (\ref{2.3}) is
likely to be due to an erroneous definition in Ref.
\cite{wonneberger}.]

\begin{figure}[htbp]
\centering
\includegraphics[width=8.8cm]{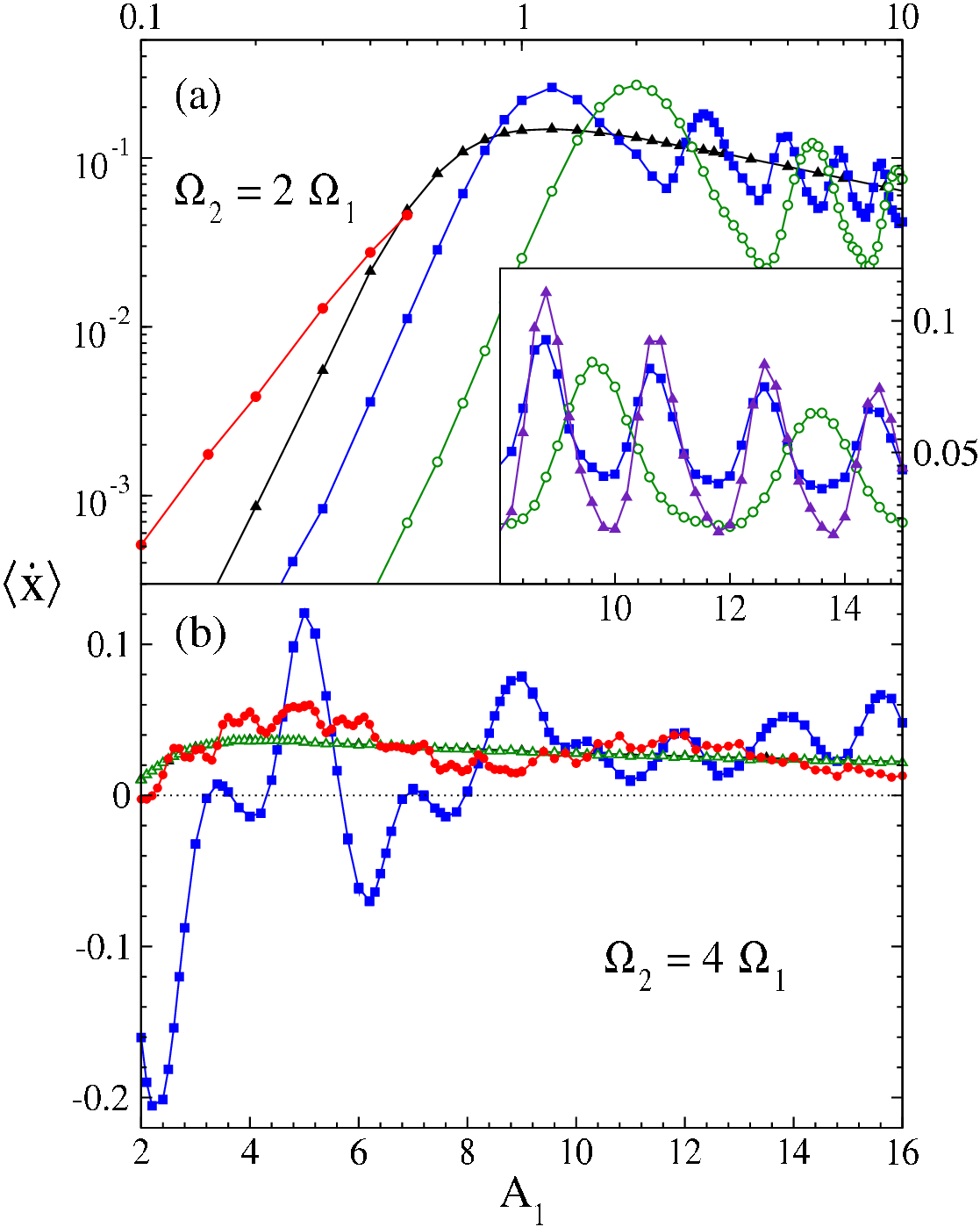}
\caption{(Color online) Transport via HM in the cosine potential (\ref{2.2}) for
$\phi_1=\phi_2$, $A_1=A_2$, and (a) $\Omega_2=2\Omega_1$, (b)
$\Omega_2=4\Omega_1$: $\langle \dot x \rangle$ versus $A_1$.
Simulation parameters: (a) squares: $\Omega_1=0.4$, $D=0.2$; empty
circles: $\Omega_1=0.8$, $D=0.2$; triangles: $\Omega_1=0.01$,
$D=0.2$; solid circles: $\Omega_1=0.05$, $D=0.4$. Inset:
squares: $\Omega_1=0.4$, $D=0.2$; empty
circles: $\Omega_1=0.8$, $D=0.2$; triangles: $\Omega_1=0.4$,
$D=0.1$; (b) squares:
$\Omega_1=0.4$, $D=0.2$; circles: $\Omega_1=0.1$, $D=0.2$;
triangles: $\Omega_1=0.01$, $D=0.2$. In both panels $d=1$.}
\label{F2}
\end{figure}

The numerical dependence of $\langle \dot x\rangle$ on the
amplitude of $F(t)$ is also more complicated than expected from
the perturbation estimate (\ref{2.3}). In Fig. \ref{F2} the time average
of $\dot x(t)$ is plotted versus $A \equiv A_1=A_2$ at
different drive frequencies $\Omega_2=2\Omega_1$. For low drive
amplitudes the HM signal $\langle \dot x\rangle$ grows indeed
proportional to $A^3$, as suggested by the scaling law (\ref{2.3}),
but only for a sufficiently high noise level $D$.

Moreover, Fig. \ref{F2} illustrates another interesting property
of rectification by HM: at relatively high ac frequencies
(non-adiabatic regime), the curves $\langle \dot x (A)\rangle$
develop regular oscillations for $A>1$ with period and amplitude
roughly proportional to $\Omega_1$. The details of such a
non-adiabatic mechanism are explained in Ref. \cite{det-rat}: On
setting $A$ at increasingly high values above the depinning
threshold of $V(x)$, max$\{|V'(x)|\}=1$, 
the number of substrate cells the driven particle crosses during
one half-cycle, increases by one unit, first to the right and then
to the left, thus causing one full $\langle \dot x\rangle$
oscillation at regular $A$ increments, $\Delta A$, proportional to
$\Omega_1$. Of course, in the adiabatic limit, $\Omega_1 \to 0$,
these oscillations tend to disappear with $\Delta A$. Moreover,
shortening the drive period or lowering the noise level for $A>1$
enhances the above modulation effect \cite{det-rat}. Finally, on
further increasing $A$ the cancellation of the right and the left
drifts becomes more and more efficient; as a result the envelope
of the $\langle \dot x\rangle$ oscillations in Fig. \ref{F2}
decays seemingly inversely proportional to $\sqrt{A}$.

\begin{figure}[htbp]
\centering
\includegraphics[width=8.8cm]{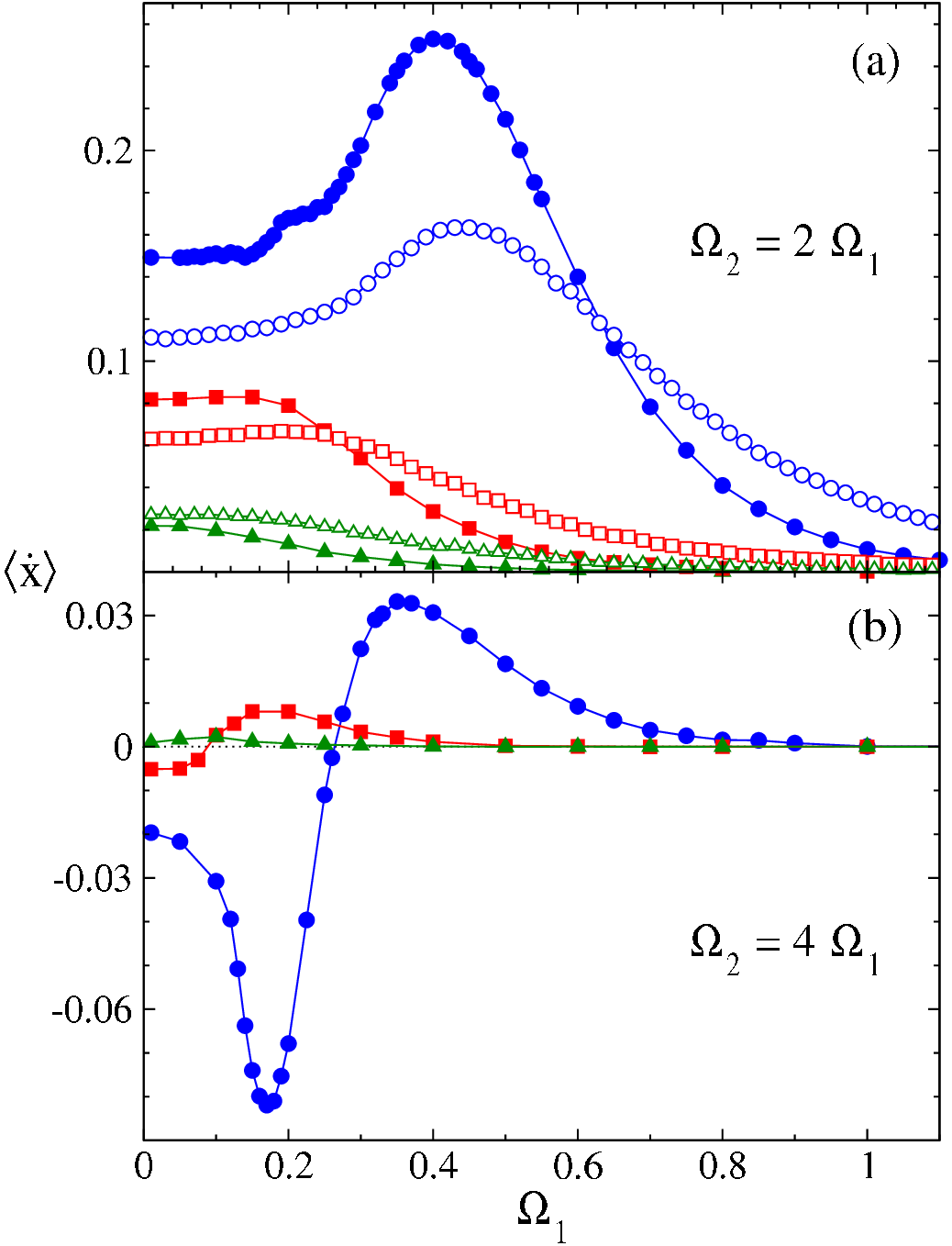}
\caption{(Color online) Transport via HM in the cosine potential (\ref{2.2}) for
$\phi_1=\phi_2$, $A_1=A_2$, and (a) $\Omega_2=2\Omega_1$, (b)
$\Omega_2=4\Omega_1$: $\langle \dot x \rangle$ versus $\Omega_1$.
Simulation parameters: solid symbols: $D=0.2$; empty symbols:
$D=0.4$; triangles: $A_1=0.4$; squares: $A_1=0.6$;  circles:
$A_1=1.1$; in both panels $d=1$.} \label{F3}
\end{figure}

An independent perturbation approach \cite{hm} led to the
following scaling law for the rectification velocity of a Brownian
particle (\ref{1.1}) in a cosine potential (\ref{2.2}) subject to
the harmonic force (\ref{2.1}) with $\Omega_2=2\Omega_1$
\begin{equation}
\label{2.4} \frac{\langle \dot x\rangle}{\Omega_1} \propto  \left
(\frac{d}{D} \right )^2 \left (\frac{A_1}{2\Omega_1} \right )^2
\frac{A_2}{2\Omega_2}.
\end{equation}
This prediction, that applies under the conditions $d \ll\Omega_1
\ll D$, reproduces at least qualitatively both the $d \to 0$
branches of Fig. \ref{F1} and the $\Omega_1 \to \infty$ tails of the
curves $\langle \dot x(\Omega_1)\rangle$ in Fig. \ref{F3}. We
leave the task of a quantitative assessment of Eq. (\ref{2.4}) for
future simulation work. Here we limit ourselves to remarking that
for large commensurate drive frequencies, i.e.,
$\Omega_1=m\Omega_0$ and $\Omega_2=n\Omega_0$ with $\Omega_0 \to
\infty$, the HM signal drops sharply to zero.

We now address two totally different drive regimes that cannot be
assimilated to the HM phenomenon introduced above. Having in mind
the drive force (\ref {2.1}), we consider the limits:\\
{\it (i)}: $\Omega_2 \to \infty$ with $\Omega_1$ and
$A_2/\Omega_2$ constant;\\
{\it (ii)}: $\Omega_+ \to \infty$ with $\Omega_-$, $A_1/\Omega_1$
and $A_2/\Omega_2$ constant.

Here we have introduced the shorthand notation $\Omega_{\pm}=
\frac{1}{2}(\Omega_1 \pm \Omega_2)$. In $(i)$ the
frequency and the amplitude of one harmonic component of $F(t)$
are taken both large, but with constant amplitude-to-frequency
ratio; in $(ii)$ the frequency and the amplitude of both
harmonic components are taken large with constant
amplitude-to-frequency ratios and, additionally, with constant
beating frequency $2\Omega_-$. The response of system (\ref{1.1})
to an external drive (\ref{2.1}) in regime $(i)$ or $(ii)$ can be
well reproduced within Bleckman's perturbation scheme, or {\it
vibrational mechanics} \cite{book}, outlined in the forthcoming
Sec. \ref{S3}. Correspondingly, the ensuing frequency mixing effect
discussed below will be termed {\it vibrational mixing}.

\section{The vibrational mechanics scheme} \label{S3}

Let us consider the overdamped Brownian particle (\ref{1.1}) with
$F(t)$ and
$\xi(t)$ defined in Eqs. (\ref{1.2}) and (\ref{1.3}),
respectively. The periodic substrate potential $V(x)$ has period
$L=2\pi$ and general form
\begin{equation}
\label{3.1} V(x) = \sum_{n=1}^{\infty} a_n \cos (nx) +
\sum_{n=1}^{\infty} b_n \sin (nx),
\end{equation} for an
appropriate choice of the Fourier coefficients $\{a_n\}$ and
$\{b_n\}$.

\begin{figure}[htbp]
\centering
\includegraphics[width=8.6cm]{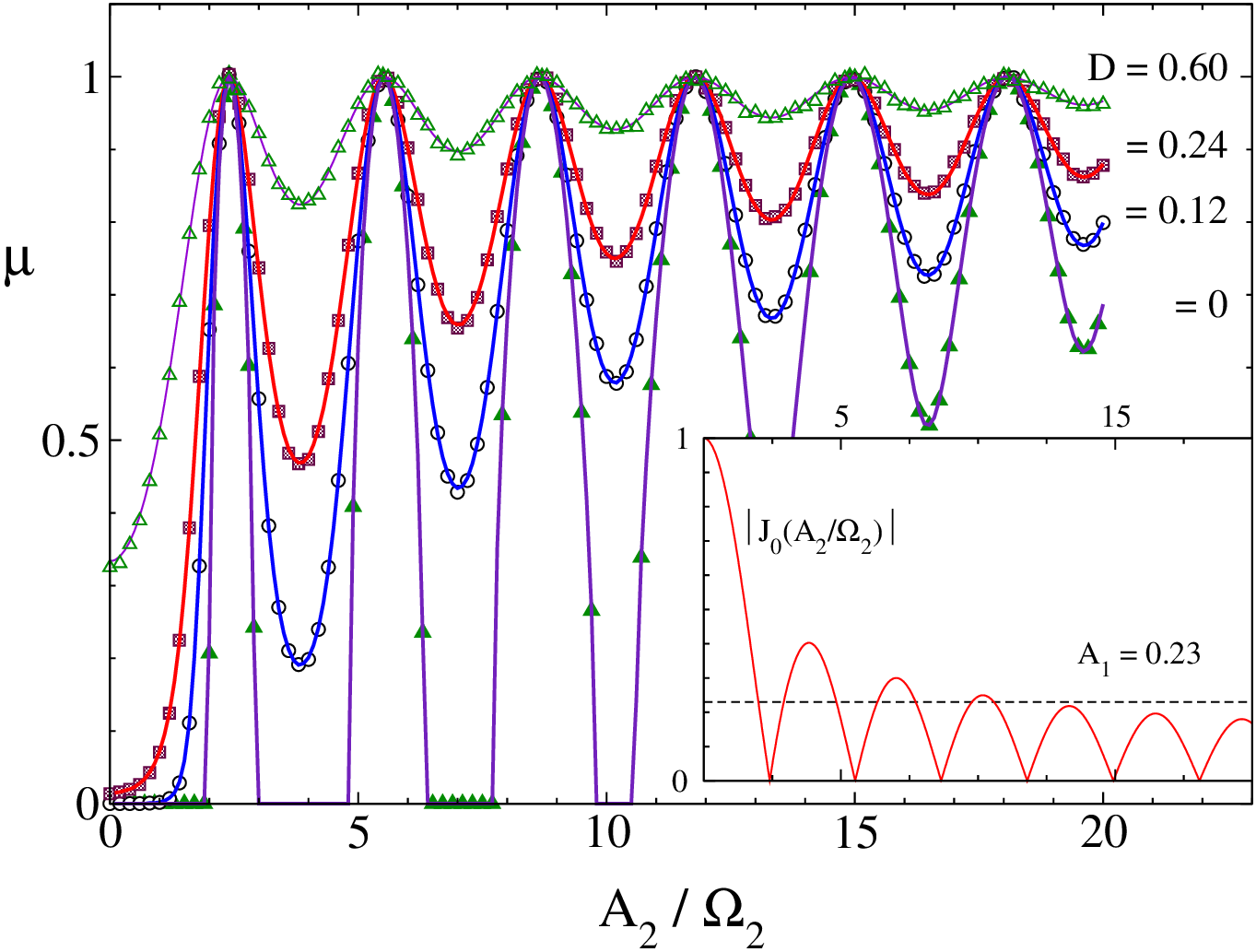}
\caption{(Color online) Mobility versus $A_2/\Omega_2$ in the dc case,
$\Omega_1=0$ and $\phi_1=0$, for different $D$.  The simulation
data (dots) have been obtained by integrating numerically the LE
(\ref{1.1}) with $V(x)$ given in Eq. (\ref{3.7}) and parameter
values: $A_1=0.23$ and $\Omega_2=0.1$. The solid curves represent
the corresponding analytic prediction (11.51) of Ref.
\cite{risken} for the reduced LE (\ref{3.4}). Inset: amplitude
$|J_0(A_2/\Omega_2)|$ of ${\overline V}(x)$ (solid curve) compared
with the static force $A_1$.} \label{F4}
\end{figure}

Let us consider for simplicity the regime $(i)$ of Sec. \ref{S2},
that is we assume that one component of $F(t)$ is slow and the
other one is fast, say, $\Omega_1 \ll \Omega_2$; then, following
the approach of Refs. \cite{landa,VR}, we can separate
\begin{equation}
\label{3.2} x(t) \longrightarrow x(t) + \psi(t),
\end{equation}
where, in shorthand notation, from now on $x(t)$ represents a
slowly time-modulated stochastic process and $\psi(t)$ is the
particle free spatial oscillation
\begin{equation}
\label{3.3}
\psi(t) = \psi_0 \sin (\Omega_2 t + \phi_2)
\end{equation}
with amplitude $\psi_0=A_2/\Omega_2$. On averaging out $\psi(t)$
over time, the LE for the slow reduced spatial variable $x(t)$ can
be written as
\begin{equation}
\label{3.4} \dot x= -{\overline V}'(x) + A_1\cos(\Omega_1 t +
\phi_1) + \xi(t),
\end{equation}
where
\begin{equation}
\label{3.5} {\overline V}(x) = \sum_{n=1}^{\infty} a_n
J_0(n\psi_0)\cos (nx) +
 \sum_{n=1}^{\infty} b_n J_0(n\psi_0) \sin (nx).
\end{equation}
Here, we made use of the identities $\langle \sin[n
\psi(t)]\rangle_t =0$ and $\langle \cos[n \psi(t)]\rangle_t =J_0(n
\psi_0)$, with $J_0(x)$ denoting the Bessel function of 0-order
\cite{bessel} -- see also inset of Fig. \ref{F4} - and $\langle (
\dots ) \rangle_t$  representing the time average of the argument $(
\dots )$.

This is an instance of the adiabatic elimination of a fast
oscillating observable \cite{grigolini},
$\psi(t)$, with {\it constant} amplitude $\psi_0$.
As a result, the slow observable $x(t)$ diffuses on an effective, or
renormalized potential ${\overline V}(x)$ driven by the slow
harmonic in Eq. (\ref{1.2}), alone. We remark that ${\overline
V}(x)$ depends on the ratio $\psi_0=A_2/\Omega_2$, the amplitude
of its $n$-th Fourier component oscillating like $|J_0(n\psi_0)|$.
The adiabatic separation (\ref{3.2}) for $\Omega_1 \ll \Omega_2$
is tenable as long as the fast oscillation amplitudes are clearly
distinguishable with respect to the corresponding Brownian
diffusion \cite{grigolini}, 
that is $\langle \psi(t)^2\rangle_t =\frac{1}{2}\psi_0^2 \gg 2Dt_2$ with
$t_2=2\pi/\Omega_2$ or, equivalently,
\begin{equation}
\label{3.6}
D \ll \frac{A_2}{8\pi}\left ( \frac{A_2}{\Omega_2}\right ).
\end{equation}
In the limit $\Omega_2 \to \infty$ at constant $A_2/\Omega_2$, the
approximate LE (\ref{3.4}) is expected to be very accurate, regardless of
the value of $D$.

We discuss now a simple application of the vibrational mechanics
scheme in the presence of a dc drive, i.e., $\Omega_1=0$ and
$\phi_1=0$. The simplest choice for the substrate potential is
\begin{equation}
\label{3.7}
V(x) = \cos x,
\end{equation}
corresponding to setting $a_1=1$ and all the remaining Fourier
coefficients $a_n, b_n$ to zero. The reduced problem
(\ref{3.4})-(\ref{3.5}) describes the Brownian diffusion in a
washboard potential with variable tilt $A_1$ \cite{risken}.

The observable that best quantifies the response of such a system
to the dc input $A_1$ is the mobility $\mu \equiv \langle \dot x
\rangle/A_1$. In Fig. \ref{F4} we compare the simulation data
for the full dynamics (\ref{1.1})-(\ref{1.3}) against the analytic
predictions for the static limit of the LE (\ref{3.4})-(\ref{3.5})
(i.e. when $\Omega_1=0$, $\phi_1=0$) at increasing ratios
$A_2/\Omega_2$ of the ac component of $F(t)$. The solid curves on
display have been obtained by computing the analytic expression
(11.51) of Ref. \cite{risken} for $\mu$. The agreement between
simulation and theory is surprisingly close even for noise
intensities above our threshold of confidence (\ref{3.6}).
\vglue 1.0truecm

The dependence of the system mobility on the high-frequency input
signal is remarkable:

(1) The curves $\mu(A_2/\Omega_2)$ exhibit apparent oscillations
with maxima at $\mu=1$. These peaks clearly correspond to values
of $\psi_0=A_2/\Omega_2$ for which the amplitude of ${\overline V}
(x) = J_0(\psi_0)\cos x$ vanishes -- see inset of Fig. \ref{F4};

(2) For $D=0$ one can apply Eq. (11.54) of Ref. \cite{risken} to
predict $\mu=\sqrt{1 -[J_0(A_2/\Omega_2)/A_1]^2}$ for $A_1 \geq
J_0(A_2/\Omega_2)$, and $\mu=0$ otherwise. As a consequence, the
mobility curve can develop a number of re-entrant peaks -- one for
each side-peak of $|J_0(A_2/\Omega_2)|$ higher than $A_1$ (three
in Fig. \ref{F4}; the peak at zero does not count);

(3) At finite noise intensities, $\mu$ is positive definite for
any $A_2/\Omega_2$; the  mobility peaks are located as in the
noiseless case, but grow less and less sharp as $D$ increases;

(4) The reduced LE (\ref{3.4})-(\ref{3.5}) holds good for $A_1=0$,
too. This implies that, for an appropriate choice of $\psi_0$, a
high-frequency sinusoidal drive $A_2 \cos (\Omega_2 t + \phi_2)$
can neutralize the effective substrate potential ${\overline
V}(x)$. Explicit numerical simulations (not shown) substantiate
this claim. For instance, the time-averaged probability density
$P(x)$ of the stochastic process (\ref{1.1}), (\ref{1.3}) and (\ref{3.7})
flattens out for $\psi_0$ approaching a zero of the Bessel
function $J_0(\psi_0)$.

Properties (1)-(4) fully establish the asymptotic regime $\Omega_2
\to \infty$ at constant $A_2/\Omega_2$ for the dynamics
(\ref{1.1}).
\vglue 1.0truecm

\begin{figure}[htbp]
\centering
\hspace*{-0.5cm} \includegraphics[width=8.8cm]{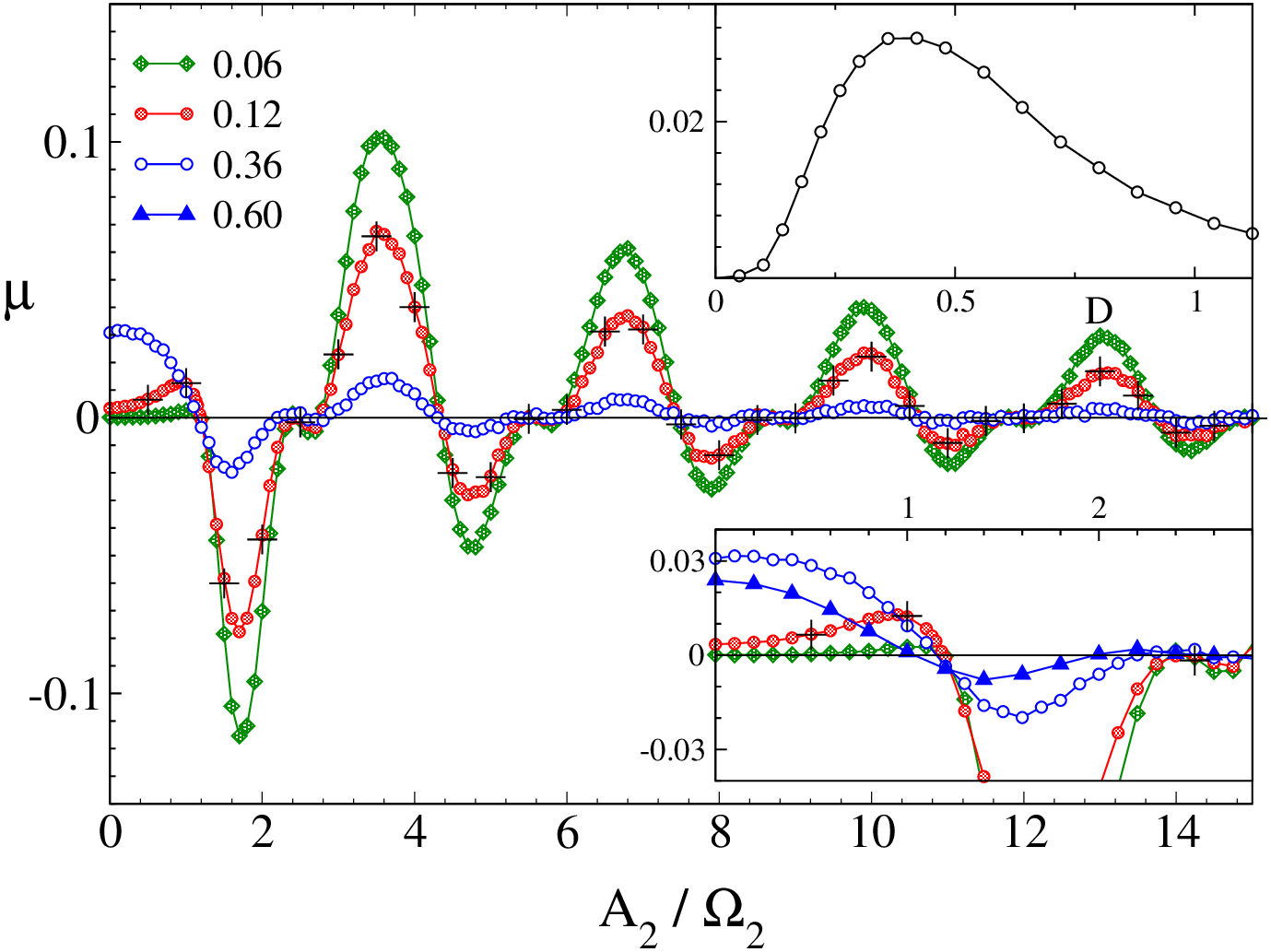}
\caption{(Color online) Mobility versus $A_2/\Omega_2$ for the doubly-rocked
ratchet (\ref{1.1}) and (\ref{4.1}) with $A_1=0.5$,
$\Omega_1=0.01$, $\phi_1=\phi_2=0$, and different values of the
noise intensity $D$. All simulation data have been obtained for
$\Omega_2=10$, but the black crosses where we set $D=0.12$ and
 $\Omega_2=20$.
Bottom inset: simulation data for $\mu(A_2/\Omega_2)$ as in the
main panel with an additional curve at $D=0.6$. Top inset: $\mu$
versus $D$ for $A_2=0$, $A_1=0.5$, and $\Omega_1=0.01$; circles:
simulation data; solid curve: adiabatic formula (11.44) of Ref.
\cite{risken}.} \label{F5}
\end{figure}

\section{Rocked vibrational ratchets} \label{S4}

We consider now a more complicated
example that falls under the category of the rocked ratchets
\cite{bartussek}. The motion of a Brownian particle on an
asymmetric substrate gets rectified when driven by a
time-correlated force, either stochastic or periodic
\cite{reimann}. Let the Fourier coefficients of the expansion
(\ref{3.1}) all be zero but $b_1=-1$ and $b_2=-\frac{1}{4}$, i.e.,
\begin{equation}
\label{4.1}
V(x) = -\sin x - \frac{1}{4}\sin 2x.
\end{equation}
The corresponding LE (\ref{1.1}) describes a doubly-rocked ratchet
\cite{riken2}. For arbitrary input frequencies $\Omega_1$,
$\Omega_2$, the rectified current of the system is known to
exhibit marked commensuration effects and a complicated dependence
on the noise intensity and all forcing parameters
\cite{riken2,chaos}. We claim here that a well-defined adiabatic limit
exists for $\Omega_1 \to 0$ and $\Omega_2 \to \infty$ with
$A_2/\Omega_2$ constant, as suggested by
the separation scheme (\ref{3.2}). Following the notation of Refs.
\cite{landa,VR}, we term a rocked ratchet operated under such
conditions a rocked {\it vibrational} ratchet (VR).

\begin{figure}[htbp]
\hspace*{-0.5cm} \includegraphics[width=8.8cm]{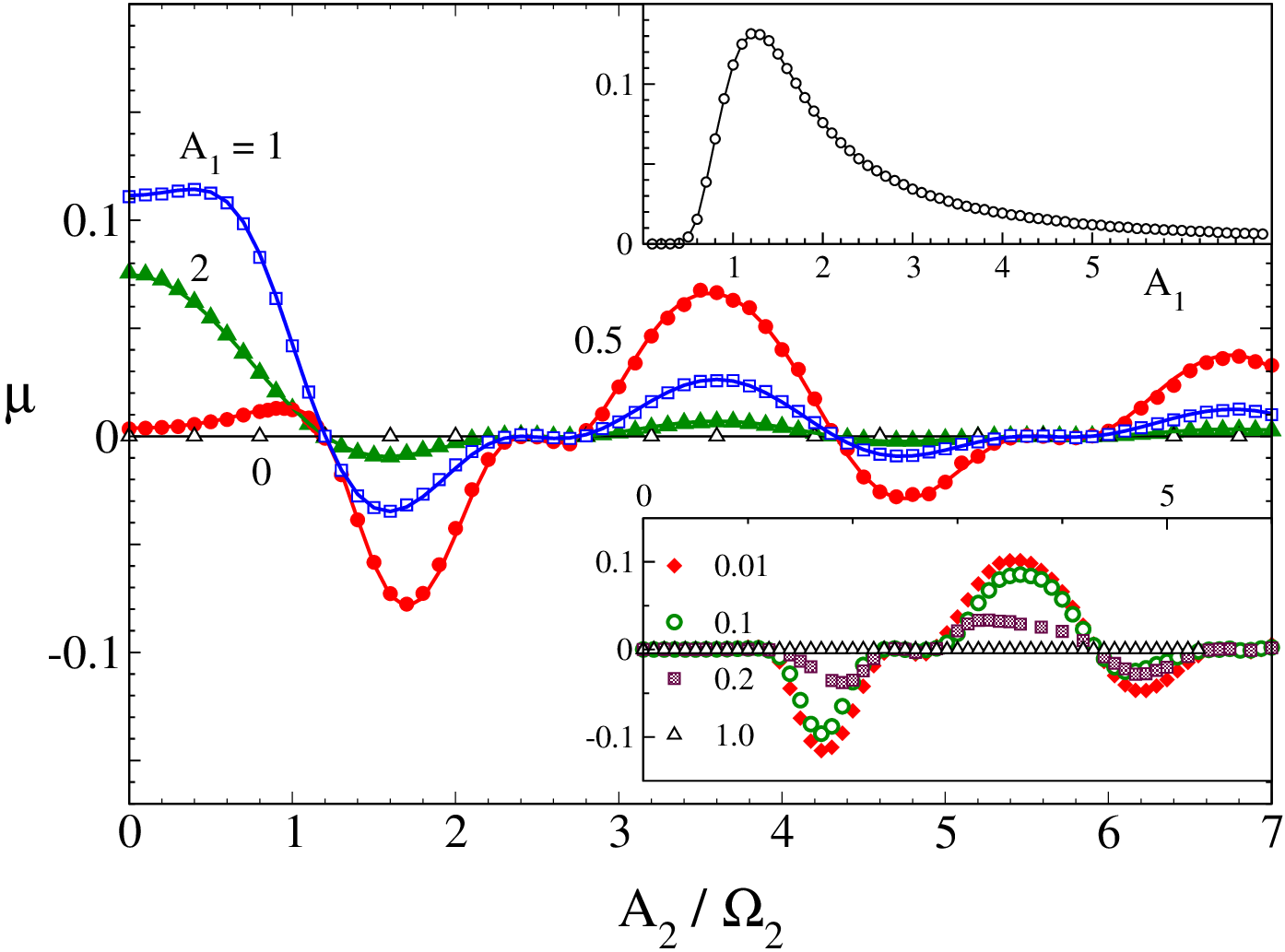}
\caption{(Color online) Mobility versus $A_2/\Omega_2$ for the doubly-rocked
ratchet (\ref{1.1}) and (\ref{4.1}) with $D=0.12$,
$\Omega_1=0.01$, $\Omega_2=10$, $\phi_1=\phi_2=0$, and different
values of $A_1$. Top inset: $\mu$ versus $A_1$ for $A_2=0$,
$D=0.12$, and $\Omega_1=0.01$; circles: simulation; solid curves:
adiabatic approximation (11.44) of Ref. \cite{risken}. Bottom
inset: $\mu$ versus $A_2/\Omega_2$ for the for the doubly-rocked
ratchet (\ref{1.1}) and (\ref{4.1}) with $A_1=0.5$, $D=0.12$,
$\Omega_2=10$, $\phi_1=\phi_2=0$, and different $\Omega_1$.
 } \label{F6}
\end{figure}

The results of our simulation work are summarized in Figs.
\ref{F5} and \ref{F6}. To explain the persistent oscillations of
the curves $\mu(A_2/\Omega_2)$, we write down explicitly the
renormalized potential i.e.
\begin{equation}
\label{4.2}
{\overline V}(x) = -J_0(\psi_0) \sin x - \frac{1}{4}
J_0(2\psi_0) \sin 2x.
\end{equation}
As long as our adiabatic elimination procedure applies, the
ratchet current $j=\langle \dot x \rangle/2\pi$ vanishes in
correspondence of the zeros of either Bessel function in Eq.
(\ref{4.2}), due to the restored symmetry of the effective
substrate. On denoting by $j_n$ the $n$-th zero of $J_0(x)$, one
predicts the following sequence of mobility zeros:
\begin{equation}
\label{4.3}
\frac{A_2}{\Omega_2} = \frac{1}{2} j_1, j_1, \frac{1}{2} j_2, \frac{1}{2} j_3, j_2,
\frac{1}{2} j_4, \frac{1}{2} j_5, j_3, \dots
\end{equation}
with $j_1=2.405$, $j_2=5.520$, $j_3=8.654$, $j_4=11.79$,
$j_5=14.93$, etc. \cite{bessel}.

As shown in Fig. \ref{F5}, the sequence (\ref{4.3}) reproduces
very closely the zero-crossings of our simulation curves for small
noise intensities; for $D=0.06$ we could locate correctly over 20
zeros of the curve $\mu(A_2/\Omega_2)$. In our derivation of the
effective potential (\ref{4.2}) we cautioned that discrepancies
may occur for $D$ above the confidence threshold (\ref{3.6}); the
deviations observed in the bottom inset of Fig. \ref{F5}
invalidate our approximation scheme only for $D \gtrsim 1$. The
amplitudes of the large $\mu(A_2/\Omega_2)$ oscillations decay
like $(A_2/\Omega_2)^{-\frac{1}{2}}$ as expected after noticing
that the modulus of $J_0(x)$ vanishes asymptotically like
$\sqrt{2/\pi x}$ for $x \to \infty$ \cite{bessel}.

Not all zeros of the sequence (\ref{4.3}) mark an inversion of the
ratchet current. For instance, for $A_2/\Omega_2 < \frac{1}{2}
j_1$ the current in the effective ratchet potential, (\ref{3.4})
and (\ref{4.2}), is certainly positive in the low $\Omega_1$
frequency regime \cite{bartussek}; for $\frac{1}{2} j_1<
A_2/\Omega_2 <j_1$, the coefficient of $\sin 2x$ changes sign and
so does the ratchet polarity (and current); on further increasing
$A_2/\Omega_2$ larger than $j_1$, the sign of both Fourier
coefficients (\ref{4.2}) get reversed with respect to (\ref{4.1}):
this is equivalent to turning $V(x)$ upside-down (beside slightly
re-modulating
 its profile), so that
the polarity of ${\overline V}(x)$ stays negative. Following this
line of reasoning one predicts {\it double} zeros (i.e. no current
inversions) at $\psi_0 = j_1, j_2, j_3, j_4, \dots$.

In the low frequency regime, $\Omega_1 \ll 1$, the reduced ratchet
dynamics, (\ref{3.4}) and (\ref{4.2}), can be treated
adiabatically. Its mobility can be computed analytically by time
averaging Eq. (11.44) of Ref. \cite{risken} over one forcing cycle
$t_1=2\pi/\Omega_1$. In Fig. \ref{F6} the analytic curves for
$\mu(A_2/\Omega_2)$ fit very closely our simulation data (dots of
the same color) at low noise, no matter what the amplitude $A_1$
of the slow harmonic in (\ref{1.2}). In the bottom inset of Fig.
\ref{F6} deviations from the low frequency curve become visible
for $\Omega_1 \gtrsim 0.1$: this does not imply that the
projection scheme leading to the reduced LE
(\ref{3.4})-(\ref{3.5}) fails on increasing $\Omega_1$ with
$\Omega_1 \ll \Omega_2$, but rather that the adiabatic treatment
of the reduced LE becomes untenable. This conclusion is
corroborated by the fact that the mobility zeros (and signs) of
the curves both in the main panel and in the bottom inset of Fig.
\ref{F6} are independent of either parameters $A_1$ and $\Omega_1$
of the low-frequency component.

Figure \ref{F6} illustrates another important VR property. 
In the presence of the high-frequency
harmonic, alone, $A_1=0$ and $\Omega_2\gg 1$, the simulated net
current is vanishingly small (triangles in the main panel).
In the absence of fast oscillations, $A_2=0$, instead, the curve
$\mu(0)$ versus $A_1$ is well reproduced by the adiabatic limit
$\Omega_1\ll 1$ \cite{bartussek} (Fig. \ref{F6}, top inset). On
comparison, one notices that, for relatively small $A_1$, the
amplitude of the $\mu(A_2/\Omega_2)$ oscillations can grow notably
larger than the corresponding $\mu(0)$. This means that energy
pumped into the system at too high frequency gets dissipated into
the heat bath, if the system is operated at equilibrium; vice
versa the nonlinear nature of the system induces a cooperative
coupling between high-frequency disturbances and optimal drives,
thus enhancing the system response beyond the expectations of the
linear response theory.

\vskip 1.0cm

\section{Pulsated vibrational ratchets} \label{S5}

\begin{figure}[htbp]
\hspace*{-0.5cm} \includegraphics[width=8.8cm]{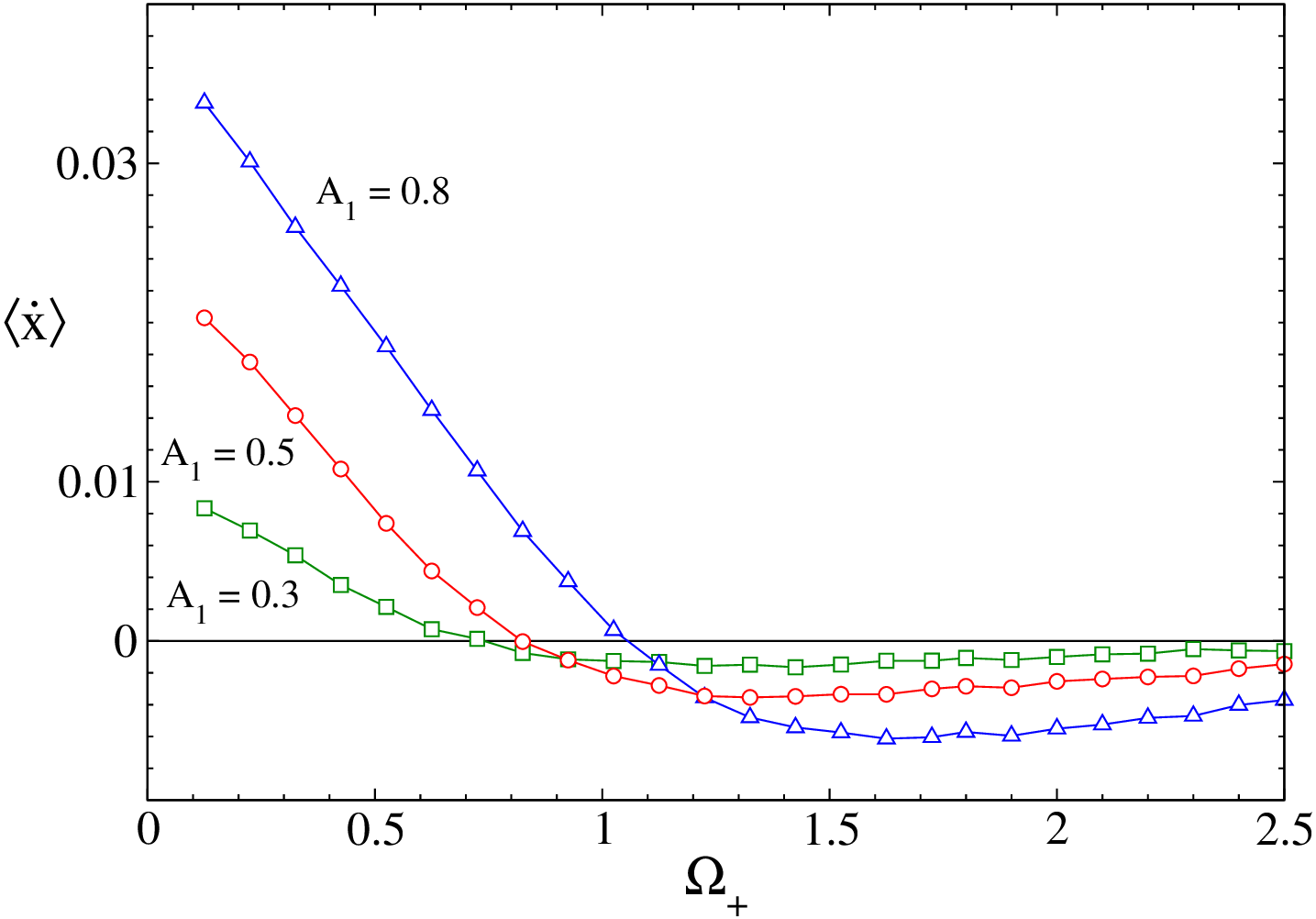}
\caption{(Color online) Transport in a VR (\ref{1.1}) and (\ref{4.1}) with
$D=0.06$, $\Omega_-=0.01$,  $\phi_1=\phi_2=0$, and different
values of $A_1=A_2$: $\langle \dot x \rangle$ versus $\Omega_+$.} \label{F7}
\end{figure}

\begin{figure}[htbp]
\hspace*{-0.5cm} \includegraphics[width=8.8cm]{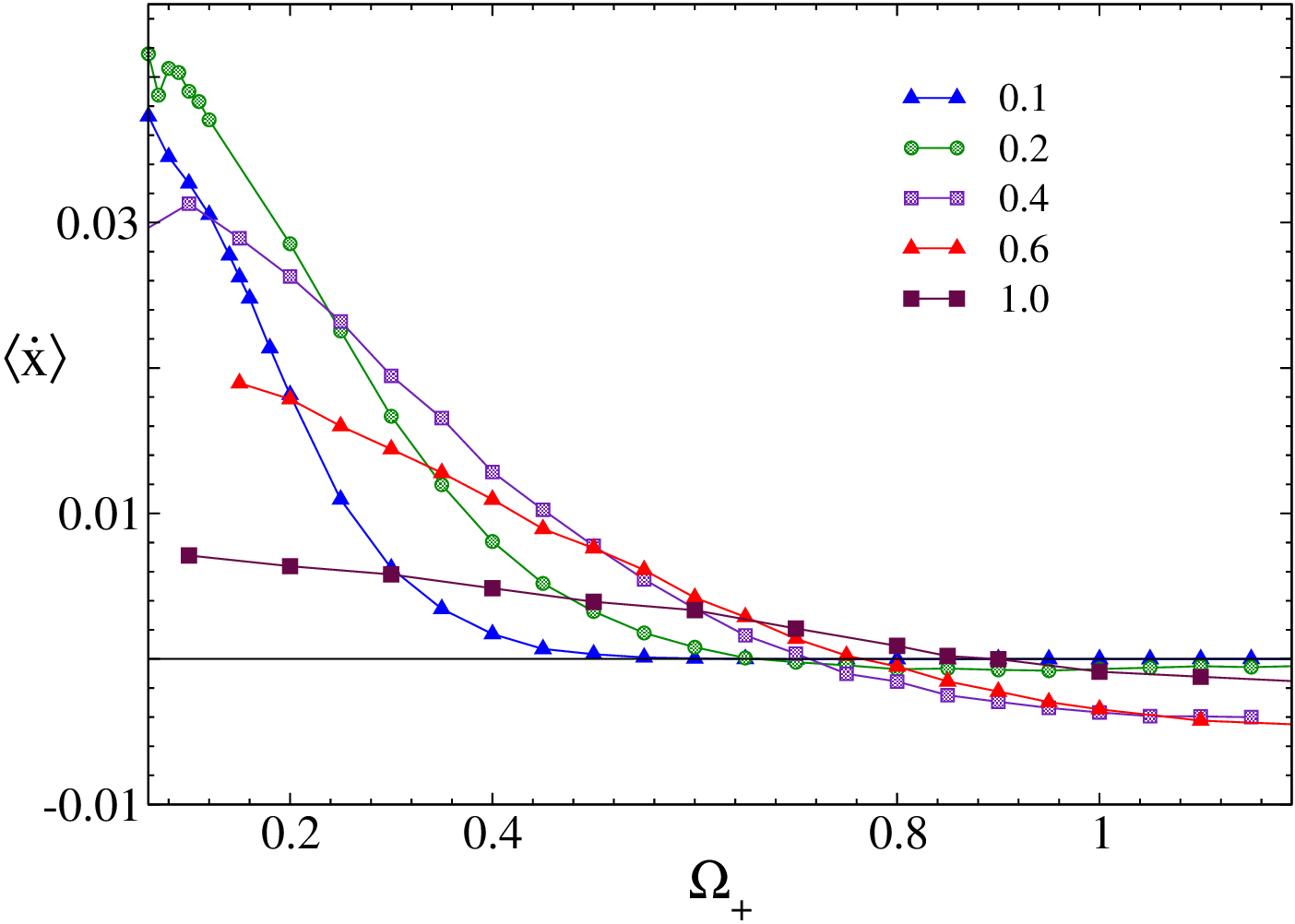}
\caption{(Color online) Transport in a VR (\ref{1.1}) and (\ref{4.1}) with
$A_1=A_2=0.5$, $\Omega_-=0.01$,  $\phi_1=\phi_2=0$, and different
values of $D$: $\langle \dot x \rangle$ versus $\Omega_+$.} \label{F8}
\end{figure}

\begin{figure}[htbp]
\hspace*{-0.5cm} \includegraphics[width=8.8cm]{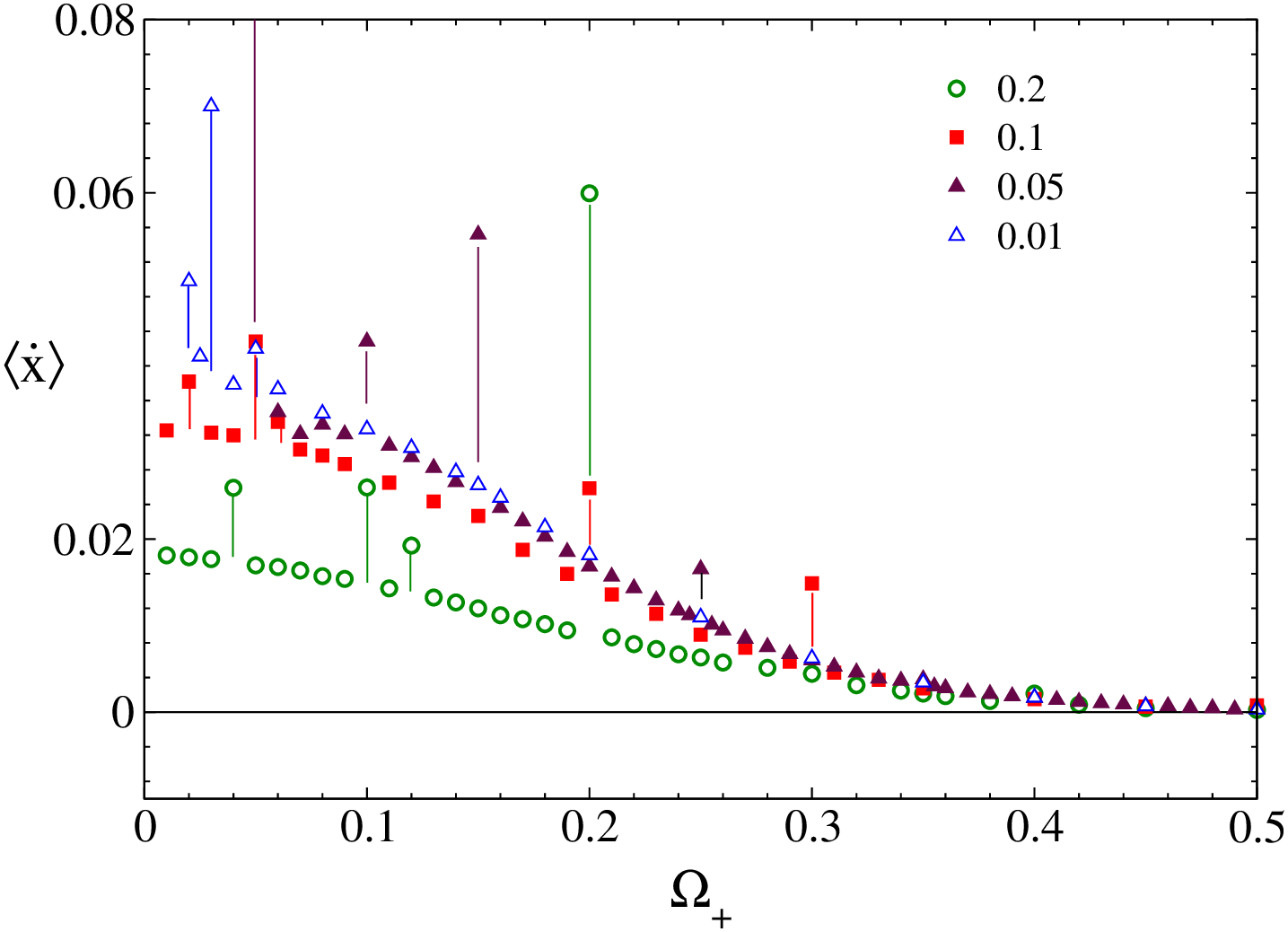}
\caption{(Color online) From harmonic to vibrational mixing: $\langle \dot x
\rangle$ versus $\Omega_+$ in the bi-harmonically rocked ratchet
(\ref{4.1}) with $A_1=A_2=0.5$, $\phi_1=\phi_2=0$, and different
values of $\Omega_-=0.01$ and $D$. Spikes appear in correspondence
with rational values of $\Omega_2/\Omega_1$, i.e. for commensurate
harmonic drives.} \label{F9}
\end{figure}

We study now the process (\ref{1.1})-(\ref{1.3}) in the regime
{\it (ii)}, i.e., for $\Omega_1, \Omega_2 \to \infty$ with
$|\Omega_2-\Omega_1|$, $\psi_1\equiv A_1/\Omega_1$, and
$\psi_2\equiv A_2/\Omega_2$ constant. In order to simplify the
algebraic passages below, we set $\phi_1=\phi_2$, like in the
simulations of Figs. \ref{F7}-\ref{F9}.

Simple trigonometric manipulations lead to the following
expression for the driven free-particle oscillations
\begin{eqnarray}
\label{5.1} \psi (t) &=& (\psi_1 +\psi_2)\sin(\Omega_+
t)\cos(\Omega_- t)\\ \nonumber &+& (\psi_1 -\psi_2)\cos(\Omega_+
t)\sin(\Omega_- t)
\end{eqnarray}
with
\begin{equation}
\label{5.2} \psi_1 \pm \psi_2 = \frac{\Omega_+(A_1\pm
A_2)-\Omega_-(A_1\mp A_2)}{\Omega_+^2-\Omega_-^2}.
\end{equation}
The parameter range relevant to the discussion of our simulation
results is
$$
\frac{A_1+A_2}{\Omega_+} \gg \left | \frac{A_1-A_2}{\Omega_-}
\right |,$$ so that
\begin{equation}
\label{5.3}
\psi_1 + \psi_2 \simeq \frac{A_1 + A_2}{\Omega_+}
\end{equation}
and
\begin{equation}
\label{5.4} \psi_1 - \psi_2 \simeq
-\frac{\Omega_-}{\Omega_+^2}(A_1 + A_2),
\end{equation}
with $\psi_1-\psi_2$ negligible with respect to $\psi_1+\psi_2$.

On applying the vibrational mechanics scheme of Sec. \ref{S3}, the
effective LE for the reduced spatial variable $x(t)$ now reads
\begin{equation}
\label{5.5}
\dot x = - {\overline V}'(x,t) + \xi(t),
\end{equation}
where
\begin{equation}
\label{5.6} {\overline V}(x,t) = \sum_{n=1}^{\infty}
J_0 \left [n\frac{A_1+A_2}{\Omega_+}\cos(\Omega_- t)\right][a_n
\cos (nx) + b_n  \sin (nx)].
\end{equation}

\vskip 1.0 cm

Here we made use of the fact that $|\Omega_-| \ll \Omega_+$, so
that the time average $\langle \cos[n\psi(t)]\rangle_t$ could be
taken over one fast oscillation cycle $T_+=2\pi/\Omega_+$, while
the slow amplitude modulation of period $T_-=2\pi/|\Omega_-|$ was
handled as an adiabatic perturbation.

The effect of the bi-harmonic drive $F(t)$ on the substrate potential
is twofold:

{(1)} The overall amplitude of the periodic function ${\overline
V}(x)$ is modulated in time, which is equivalent to periodically
modulating the noise intensity $D$. This condition is reminiscent
of the so-called {\it temperature} ratchets \cite{temp-rat}
discussed at length in Ref. \cite{reimann};

{(2)} Since the argument of the Bessel functions in Eq.
(\ref{5.6}) depends on the index $n$, the Fourier coefficients of
${\overline V}(x)$ are distinctly modulated in time;
as a consequence, their relative weights change in time and so
does the profile of ${\overline V}(x)$.

Separating (sure, rather arbitrarily!) the two time dependencies
{(1)} and {(2)} of ${\overline V}(x)$ helps us explain the
simulation results of Figs. \ref{F7}-\ref{F9} obtained for the
asymmetric potential (\ref{4.1}).

\vskip 1.0cm

Figures \ref{F7} and \ref{F8} illustrate the most interesting
feature of this class of VR, namely the inversion from positive to
negative current that takes place for $\Omega_+ \gg |\Omega_-|.$
We recall that the positive orientation of $\langle \dot x\rangle$
corresponds to the normal polarity of a ratchet (\ref{4.1}) slowly
rocked by a harmonic drive, whereas the negative orientation is to
be expected when the same device is being operated in the pulsated
mode (like for an adiabatic temperature ratchet \cite{temp-rat}).
This is why we term regime {\it (ii)} of the process
(\ref{1.1})-(\ref{1.3}) {\it pulsated} VR.

Note that the pulsated VR of  Figs.  \ref{F7} and \ref{F8} are
operated for values of $\Omega_-$ and $\Omega_+$ such that the HM
mechanism plays no significant role, being the HM spikes confined
to values of $\Omega_+$ much closer to $|\Omega_-|$ (as shown in
Fig. \ref{F9}).

The current inversions in the adiabatic regime $\Omega_+ \gg |\Omega_-|$
can be explained by looking at the time-dependent effective potential
\begin{eqnarray}
\label{5.7} {\overline V}(x,t) = &-& J_0 \left
[\frac{2A}{\Omega_+}\cos(\Omega_- t)\right] \sin (x)\\ \nonumber
 &-& \frac{1}{4} J_0 \left [\frac{4A}{\Omega_+}\cos(\Omega_-
t)\right] \sin (2x),
\end{eqnarray}
associated with the substrate potential $V(x)$ of Eq. (\ref{4.1}).
Here we set $\phi_1=\phi_2$ and $A_1=A_2 \equiv A$ to make contact
with the simulation conditions of Figs. \ref{F7} and \ref{F8}. For
$|4A/\Omega_+|<j_1$, i.e., for $\Omega_+ > \Omega_+^*$ with
$\Omega_+^*=4A/j_1$, both Fourier coefficients in Eq. (\ref{5.7})
retain their (negative) sign at any time $t$; ${\overline V}(x,t)$
does not change polarity and the overall effect of the adiabatic
modulation with period $T_-$ amounts to pulsating  the amplitude
of the effective potential (or, equivalently, the noise level 
\cite{temp-rat})
with the same period. As a consequence, $\langle \dot x\rangle$ is
predicted to change sign from positive for $\Omega_+ < \Omega_+^*$
to negative for $\Omega_+ > \Omega_+^*$. No more current reversals
are expected for higher $\Omega_+$; as usual in the ratchet
phenomenology, $\langle \dot x\rangle$ tends to zero for $\Omega_+
\to \infty$.

In the actual simulation, see e.g. Figs. \ref{F7} and \ref{F8}, these current
inversions seem to take place for $\Omega_+$ slightly smaller than
$\Omega_+^*$. This is due to the fact that at $\Omega_+ =
\Omega_+^*$ the second coefficient of Eq. (\ref{5.7}) is 
negative over the entire averaging cycle $T_-$, but for $t=mT_-$
with $m=0,1,2 \dots$, where it vanishes. 
This means that the average $\langle \dot
x\rangle$ is still negative at $\Omega_+ = \Omega_+^*$ and
vanishes only for lower (but not too lower)
$\Omega_+$ values -- in strict sense, $\Omega_+^*$ is only an
upper bound to the cross-over frequency. This
argument applies as long as condition (\ref{3.6}) holds; that is not the
case of $A_1=0.3$ in Fig. \ref{F7} and $D=1.0$ in Fig. \ref{F8}.

Finally, we note that the dependence of $\langle \dot x\rangle$
on the noise intensity $D$ exhibits the resonant behavior peculiar
to most ratchet currents \cite{reimann}.
This happens both for positive and negative rectification currents
(see Fig. \ref{F8}).

\vskip 1.0 cm

\section{Concluding remarks} \label{S6}

The robustness of the effects simulated here hints at the
possibility of implementing this concept in the design and
operation of efficient e.m. wave sensors. As a matter of fact, the
present investigation has been inspired by a typical signal
detection problem \cite{signal}, namely how to reveal a high
frequency signal by means of a sensor with optimal sensitivity in
a relatively low frequency band. All cases discussed in the
foregoing Secs. \ref{S3}-\ref{S5} suggest a simple recipe:
Although the unknown high-frequency signal alone cannot be
detected, adding a tunable control signal with parameters within
the device sensitivity range causes a nonlinear transfer of energy
(information) from high to low frequencies, thus
enhancing/modulating the sensor response to the control signal. By
analyzing the dependence of the device output on the tunable input
signal, we can reveal the existence of unknown (and otherwise
undetectable) high-frequency signals. Note that in the simulations
of Figs. \ref{F5} and \ref{F6} the input frequencies $\Omega_1$
and $\Omega_2$ differ by {\it four orders} of magnitude or more,
whereas the response function $\mu$ can be sensitive to forced
output oscillation amplitudes $\psi_0$ as small as the device
substrate unit length $L$. In the simulations of Figs. \ref{F7}
and \ref{F8}, instead, the ratchet currents are controlled by
beating frequencies  $|\Omega_2-\Omega_1|$ two (or more) orders of
magnitude smaller than the carrier frequency
$\frac{1}{2}(\Omega_1+\Omega_2)$.

As a further application we suggest that the mechanism of
frequency coupling studied here can impact our assessment of the
health hazards associated with electro-pollution
\cite{pollution1}. While high-frequency (non-ionizing) e.m.
radiation is likely to be harmless at the small length scales of
sensitive biomolecules, like the DNA helix \cite{pollution2},
still it can affect physiological processes at the cell level. Not
only high-frequency e.m. waves heat up the biological tissues, but
in view the present report, such radiation can interfere with the
much lower-frequency electro-chemical control signals that
regulate the ion transport across cell membranes or the
information transfer and processing through neuron networks and
sensory nerves. In conclusion, our results corroborate the recent
shift of the biomedical research focus from structural (and
irreversible) to functional (and possibly reversible) biological
damages \cite{cell} caused by electro-pollution.

Finally, we mention another potential extension of the
vibrational mechanics scheme of Sec. \ref{S3}. In a forthcoming
paper we study the inertial effects of the LE
\begin{equation}
\label{6.1}
\ddot x = - \gamma \dot x - V'(x) +F(t) + \xi(t)
\end{equation}
with $F(t)$ and $\xi(t)$ defined in Sec. \ref{S1} and $D\equiv
\gamma kT$. Here we limit ourselves to anticipating that the vibrational
mechanics scheme of Sec. \ref{S2} still applies
upon replacing $\psi_0=A_2/\Omega_2$ with
\begin{equation}
\label{6.2}
\psi_0 \longrightarrow \psi_{\gamma} = \frac{\psi_0}{\gamma^2+\Omega_2^2}.
\end{equation}
As a consequence, $\langle \dot x\rangle$ depends on both $\psi_0$
and $\Omega_2$, which suggests the design of highly sensitive
devices capable of separating the components of a multi-species
mixture according to the different particle masses.

\end{document}